\def\virg#1{``#1''}

\def\rfr#1{(\ref{#1})}

\def\bm#1{{\mbox{\boldmath$#1$\unboldmath}}}

\def\bar{\begin{eqnarray}}
\def\ear{\end{eqnarray}}

\def\eqi{\begin{equation}}
\def\eqf{\end{equation}}
\def\eqia{\begin{eqnarray}}
\def\eqfa{\end{eqnarray}}
\def\rp#1#2{{#1\over#2}}

\def\lb#1{\label{#1}}

\def\dr{\bm A_{\rm drag}}

\def\oc2{$\mathcal{O}(c^{-2})$}

\documentclass[CEJP,PDF]{cej} 
\usepackage{layout}
\usepackage{amsmath}
\usepackage{textcomp}

 \title{On the Lense-Thirring test with the Mars Global Surveyor in the gravitational field of Mars}

\articletype{Research article}

\author{Lorenzo~Iorio\inst{1}\email{lorenzo.iorio@libero.it},
        }

\institute{
     \inst{1} INFN-Sezione di Pisa,\\
     Viale Unit\`{a} di Italia 68, 70125 Bari, Italy
     }

\abstract{I discuss some
aspects of the recent  test of frame-dragging performed by me by exploiting the Root-Mean-Square (RMS) orbit overlap differences of the out-of-plane component $N$ of the orbit of the Mars Global Surveyor (MGS) spacecraft in the gravitational
field of Mars.    A linear fit of the full time series of the entire MGS data
(4 February 1999--14 January 2005) yields a normalized slope
$1.03\pm 0.41$ (with 95$\%$ confidence bounds). Other linear fits to different data sets confirm the agreement with general relativity.   The  huge systematic effects induced by the mismodeling in the martian
gravitational field claimed by some authors are  absent in the MGS out-of-plane record. The non-gravitational forces affect  at the same level of the gravitomagnetic one the in-plane orbital components of MGS, not the out-of-plane one. Moreover, they experience high-frequency variations which does not matter in the present case in which secular effects are relevant.
    }


\begin{document}
\maketitle

\section{Introduction}
I  proposed in Refs. \cite{Ior06,Ior07} an interpretation of the time
series of the RMS orbit overlap differences \cite{Kon06} of the out-of-plane part $N$
of the orbit of the Martian polar artificial satellite Mars Global Surveyor
(MGS) over a time span $\Delta P$ of about 5 years (14 November
1999-14 January 2005 in Ref. \cite{Ior07})
in terms of the general relativistic
gravitomagnetic Lense-Thirring effect\footnote{It consists of a small precession of the orbit of a test particle freely moving around a central rotating mass; the general relativistic component of the gravitational field due to the mass currents generated by the rotation of body is named \virg{gravitomagnetic} by analogy with the magnetic field induced by electric currents in the Maxwellian electromagnetism \cite{Mash}. For some recent, controversial tests performed in the gravitational field of the Earth with the LAGEOS satellites, see Ref. \cite{SSR} and references therein.} \cite{LT}. It turned out that the average of such a time series over
$\Delta P$, normalized to the predicted Lense-Thirring
out-of-plane mean shift over the same time span, is $\mu=
1.0018\pm 0.0053$.

Our interpretation has recently been questioned by Krogh in Ref. \cite{Kro07}.
The remarks concerning the analysis presented in Refs. \cite{Ior06,Ior07}
mainly deal with I) The observable used: in Refs. \cite{Ior06,Ior07} I would have
misinterpreted the MGS data II) The confrontation between the
prediction of the gravitomagnetic Lense-Thirring shift and the
data over the chosen time span $\Delta P$: in Refs. \cite{Ior06,Ior07} I would have
incorrectly compared the 1.6 m value of the out-of-plane average
orbit error released by Konopliv et al. in Ref. \cite{Kon06} for the entire MGS
data set to the Lense-Thirring
shift calculated for a shorter time interval $\Delta P$  III) The data set used: in Refs. \cite{Ior06,Ior07} I discarded some of the initial months of the MGS data set
 IV) The bias--neglected by me in Refs. \cite{Ior06,Ior07}--due to the multipolar expansion of the
Newtonian part of the martian gravity field, as pointed out in the unpublished Refs. \cite{Sin07,Fel07}
quoted by Krogh in Ref. \cite{Kro07} V) The impact of the atmospheric drag, neglected by me in Refs. \cite{Ior06,Ior07}

Below I present my reply which, basically, consists of the following points. As further, independent tests,  here I present various linear fits to {\it different data sets} including, among others, the {\it full} time series of the {\it entire MGS data}  (4 February 1999--14 January 2005) as well; the predictions of general relativity turn out to be {\it always confirmed}. The analytical calculation of the competing aliasing effects due to both the gravitational and non-gravitational perturbations, which affect the {\it in-plane} orbital components of MGS, do {\it not} show up in the real data.
Moreover, the non-conservative forces, whose steadily refined modeling mainly improved the {\it in-plane} orbital components of MGS, {\it not} the {\it normal} one, exhibit high-frequency, non-cumulative in time variations.

%
\section{My arguments}
\subsection{A consideration of general nature}
Before going into the details of my reply, I will extend to the MGS-Mars system a standard argument used to evaluate the possibility that general relativity has to be taken into account in describing the motion of test bodies in a given gravitational field \cite{Sof03}.  It consists of considering the relativistic characteristic length of the problem at hand, like the Schwarzschild radius of the body acting as source of the gravitational field, and comparing it to the accuracy with which the orbit of a test particle orbiting it can be determined. In the present case, the characteristic gravitomagnetic length of a rotating body of mass $M$ and angular momentum $S$ is \cite{Tarta}
\eqi l_g=\rp{S}{cM},\eqf where $c$ is the speed of light in vacuum; $l_g$ is also one of the two parameters with dimension of lengths entering the Kerr metric \cite{Kerr} which describes the exterior field of a rotating black hole.
Since the angular momentum of Mars can be evaluated as
 \eqi S_{\rm M}=(1.92\pm 0.01)\times 10^{32}\ {\rm kg\ m}^2\ {\rm s}^{-1}\eqf from the latest spacecraft-based determinations of the areophysical parameters quoted in Ref. \cite{Kon06},
 it turns out
 \eqi l^{\rm M}_g = \rp{S_{\rm M}}{cM_{\rm M}} =1.0\ {\rm m}.\eqf Such a value has to be compared with the present-day accuracy in determining the orbit of MGS; it can be evaluated it is about  $0.15$ m \cite{Kon06} in the radial direction, not affected by the gravitomagnetic force itself, as I will show later.
 Thus, it makes sense to investigate the possibility of measuring the Lense-Thirring effect in the gravitational field of Mars with MGS in such a way that a possible positive outcome should not be regarded as unlikely.
\subsection{Reply to Krogh's points}
\begin{itemize}

                    \item  [1)]

                    The entire MGS data  set  was subdivided by Konopliv et al. in Ref. \cite{Kon06} in 388 (not 442, as claimed by Krogh in Ref. \cite{Kro07}) smaller time intervals
  of data called arcs. For MGS, the lengths of the arcs vary from 4 to
6 days, so to cover many orbital revolutions ($\approx 2$ h).
  For each arc, the spacecraft position and the velocity, among other things, were estimated and used as
  starting point for a numerical propagation of the satellite's motion by
  means of the dynamical models which, in the case of MGS, did not include the general relativistic
  gravitomagnetic force.
  Contiguous arcs were overlapped by an amount of just 2 h, i.e. one
  orbital revolution, and the\footnote{I acknowledge the use of  wrong terminology in \cite{Ior06}, as noted by Kroght in Ref. \cite{Kro07}, although I feel that the complaint of plagiarism that seems to be raised by Krogh in Ref. \cite{Kro07}  in his footnote 1, pag. 5710 against me \cite{Ior07} about such a matter sounds a bit excessive.} RMS spacecraft position difference
  among the predicted positions propagated from the estimated ones in the previous arc
  and the estimated positions of the subsequent arc was computed. Since the arc overlaps cover just about one orbit,
  such RMS differences, in fact, account for any among measurement errors, random errors, systematic bias due to mismodeling/unmodeling dynamical forces
  yielding secular, i.e. averaged over one orbital
  revolution, effects, whatever their physical origin may be.
  Indeed, RMS of orbit solution overlaps are {\it commonly used in satellite
  geodesy as useful and significant indicators of the
  overall orbit accuracy} \cite{Tap04,Luc06}.
  Conversely, they are also used to gain
  information about {\it systematic errors coming from inaccurate
  modeling of the forces acting on the spacecraft}. For details
  see Refs. \cite{Tap04,Luc06}.
  Of course, such a technique is insensitive to
  short-period effects, i.e. having frequencies
  higher than the orbital one: only dynamical features of
  motion with time-scales equal to, or larger than one orbital period
  can be sensed by such orbit overlap differences. Moreover, the
  average orbit error $\left\langle\Delta N_{\rm
diff}\right\rangle$ of about 1.6 m does not refer to this or that
  particular arc overlap; instead, it comes from the mean
  of the entire set of RMS orbit overlap differences for the chosen time span $\Delta
  P$ and is well representative of those un-modelled/mis-modelled
  forces yielding effects which do not average out over $\Delta
  P$, as it is just the case of the Lense-Thirring signal.
  Time-varying patterns exhibiting well-defined periodicities-including  also measurement errors like, e.g., those related to
   the Earth-Mars geometry-are, instead,
  mainly averaged out yielding little or no contribution to
  the average orbit error. Incidentally, from the above discussion about
  the meaning of the average orbit error, it should be apparent that it does
  not make sense to look for the error of the error, as, instead, seemingly
  required by Krogh in Ref. \cite{Kro07} when he blames me \cite{Ior07} for not having included the uncertainty in $\left\langle\Delta N_{\rm
res}\right\rangle$. Another criticism by Krogh in Ref. \cite{Kro07} is that the
RMS overlap differences would be unable to specify any orbital precession.

To reply to all such criticisms I decide to perform another, independent test of my hypothesis.
First, by linearly
fitting\footnote{Note that, since the plots in Fig. 3 of Ref. \cite{Kon06} are
{\it semi-logarithmic}, one should {\it not} visually look for a straight line
in them.} the {\it full time series}  of Ref. \cite{Kon06}, after having rescaled the data points in order to shift the zero point of the time-series to the middle of the data span, I get a slope of $-0.64\pm 0.26$ m yr$^{-1}$, (with 95$\%$ confidence bounds), while the predicted
Lense-Thirring MGS out-of-plane rate (customarily defined {\it positive}
{\it along} the spacecraft's orbital angular momentum) amounts to 0.62 m yr$^{-1}$.
The obtained minus sign is due to the fact that Konopliv et al. in Ref. \cite{Kon06}
defined the normal direction to be positive in the {\it opposite} direction of
the MGS orbital angular momentum (Konopliv 2007, private communication). Should such a linear fit be used
as indicator of the existence of the Lense-Thirring effect, its relativistic prediction would be fully confirmed
within the experimental error; instead,
the hypothesis of a null effect would be rejected at 2.4 sigma level.
Then, I also repeat my procedure by fitting with a straight line the entire data set {\it without full January 2001}, mainly affected by likely measurement errors which, according to Krogh in Ref. \cite{Kro07}, would mimic the Lense-Thirring effect, getting  $-0.61\pm 0.26$ m yr$^{-1}$.  {\it The removal of the entire year 2001}, mainly affected by angular momentum wheel desaturation operations, yields $-0.57\pm 0.28$ m yr$^{-1}$. Another linear fit to the time series after removing the last month (December 2004-January 2005) yields $-0.62\pm 0.27$ m yr$^{-1}$.

Such results reply to the criticisms II) and III) as well concerning $\Delta P$, to which a large part of Ref. \cite{Kro07} is devoted.

             \item [IV)] Krogh in Ref. \cite{Kro07} quotes  Ref. \cite{Sin07}  in which analytical calculation about the corrupting impact of various physical parameters of Mars through the classical node precessions induced by the even zonal harmonic coefficients $J_{\ell}$ of the multipolar expansion of the Newtonian part of the martian gravitational potential are presented.   In particular,
                 Sindoni et al. in Ref. \cite{Sin07}  use the first five even zonals $J_2...J_{10}$ along with their associated errors from former global solutions for the Mars'gravity field, the uncertainty in the Mars'  $GM$ and in the MGS semimajor axis and inclination, plug them into analytical formulas for the classical secular node precessions and conclude that, since the resulting effect is tens of thousand times larger than the Lense-Thirring effect on MGS, this would be fatal for any attempt to detect the gravitomagnetic frame-dragging with such a spacecraft.

                 The point is that {\it such figures}, as others which can be obtained from more accurate calculation, {\it must ultimately be compared with the reality of the data}, i.e. the RMS orbit overlap differences of MGS.

                 I, in fact, repeated such calculation by considering also the other even zonals up to $J_{20}$ along with the latest errors of the  MGS95J global solution and including the uncertainties in the Mars' radius as well. By summing, in a root-sum-square fashion, such  terms I get a mean bias of 78.9 m d$^{-1}$ in the out-of-plane MGS orbital component: by linearly summing them I get an upper bound of $111.6$  m d$^{-1}$.
                 Such figures clearly show how they are by far not representative of the real MGS orbit. Indeed, over a time span of 5 years I would have an {\it enormous mean shift as large as 144 km} (root-sum-square calculation) or {\it 203 km} (linear sum). Interestingly, even if the set of the RMS overlap differences of MGS were to be considered as representative of a single orbital arc 6 d long only, the conclusion would be the same: indeed, in {\it this} case, the total cross-track mean shift due to the martian gravitational potential {\it would amount to 473.1 m} (root-sum-square) or {\it 669.6 m} (linear sum).

                 In regard to Ref. \cite{Fel07}, quoted in Ref. \cite{Kro07} as well, let us recall again that the RMS orbit overlap differences are just used to account, in general, for all the measuremnt/systematic errors giving an indication of the overall orbit accuracy
                 \cite{Tap04,Luc06}. The important point is that {\it they cancel out, by construction,} errors, systematic or not, {\it common to consecutive arcs}$-$it would just be the case of a bias like that described by Felici in Ref. \cite{Fel07}$-$, while effects like the Lense-Thirring one, accumulating in time, are, instead, singled out \cite{Luc06}.

                  \item [V)] In regard to the impact of the non-gravitational perturbations,    Sindoni et al. in Ref. \cite{Sin07}
yield a total un-modelled  non-gravitational acceleration of
$\approx 10^{-11}$ m s$^{-2}$ which is the same order of magnitude
of the Lense-Thirring acceleration induced by Mars on MGS. They  neither present any detailed calculation of the effect of such an
acceleration on the normal portion of the MGS orbit nor specify if such a magnitude refers to the out-of-plane component. However, some
simple considerations can be easily traced: a hypothetic, generic perturbing out-of-plane
force 6.7 times larger than the Lense-Thirring one and having the same time signature, i.e. linear in time, should
induce a 10.8 m cross-track shift, on average, over the considered
time span $\Delta P$. Again, {\it such a bias is neatly absent from the data}.
 By the way, as clearly stated in Ref. \cite{Kon06},
it is the {\it along-track} portion of the MGS orbit$-$left unaffected
by the Lense-Thirring force$-${\it to be mainly perturbed by the
non-gravitational forces}: indeed, the along-track empirical
accelerations fitted by Konopliv et al. in Ref. \cite{Kon06} amount just to
$\approx 10^{-11}$ m s$^{-2}$, which shows that the guess by Sindoni et al. in Ref. \cite{Sin07} is somewhat correct, but it refers to the {\it along-track} component.

Time-dependent, periodic signatures would, instead, be averaged
out, provided that their characteristic time scales are relatively short, as it is just the case. Indeed, the non-conservative accelerations, which are especially active in the MGS {\it in-plane} orbital components as clearly stated in Refs. \cite{Lem01,Kon06}, exhibit time-varying patterns over 12 hr \cite{Lem01} which, hypothetically mapped to the out-of-plane direction, are averaged
out over  multi-year time spans  (and, incidentally, over 6 d as well).
To be more definite,
in regard to the issue of the impact of the atmospheric drag on the
  cross-track portion of the orbit of MGS, raised by Krogh in Ref. \cite{Kro07}, let us note that
  it requires not only  to consider the node $\Omega$, as apparently claimed by Krogh in Ref. \cite{Kro07}, but also the inclination $i$ according to Ref. \cite{Chr88}
  \eqi\Delta N=a\sqrt{\left(1+\rp{e^2}{2}\right)\left[\rp{(\Delta i)^2}{2} + (\sin i\Delta\Omega)^2\right]}.\lb{tra}\eqf
  According to, e.g., Milani et al. in Ref. \cite{Mil87}, the perturbing acceleration $\dr$ due to the atmospheric drag
  can be cast into the form
  \eqi \dr=-\rp{1}{2}ZC_{\rm D}\rp{S}{M}\rho v\bm v,\eqf
  where $S/M$ is the spacecraft cross sectional area (perpendicular to the
  velocity) divided by its mass, $C_{\rm D}$ is the drag
  coefficient, $\rho$ is the atmospheric  density (assumed to be constant over one orbital
  revolution), $\bm v$ is the satellite velocity in a
  planetocentric, non-rotating frame of reference and $Z$ is a
  corrective coefficient accounting for the fact  that the
  atmosphere is not at rest, but rotates with angular velocity $\omega_{\rm A}$
  more or less rigidly with
  the planet; $Z\approx 1$ for polar orbits \cite{Mil87}.
  While the secular, i.e. averaged over one orbital
  period $T$, drag shift on the node vanishes, it is not so for
  the inclination: indeed, it turns out \cite{Mil87}
  \eqi \left\langle\Delta i\right\rangle_{T} \approx \pi\left(\rp{A_{\rm drag}}{n^2 a}\right)\rp{\omega_{\rm A}}{n}
  +{\mathcal{O}}(e), \eqf where $n=\sqrt{GM/a^3}$ is the Keplerian
  mean motion. As a result, the orbital plane tends to approach
  the planet's equator; the terms in brackets is the ratio of the
  drag force to the Newtonian monopole. As usual in perturbation theory, $a$ is meant as evaluated on the unperturbed reference
  ellipse. Thus, the out-of-plane drag shift
  is from \rfr{tra}
  \eqi \left\langle\Delta N_{\rm drag}\right\rangle\approx a\rp{\left\langle\Delta i_{\rm drag}\right\rangle}{\sqrt{2}}.\eqf
  In the following I will assume that $\omega_{\rm
A}\approx\omega_{\rm Mars}=7.10\times 10^{-5}$ s$^{-1}$.
  Let us see what happens in the (unlikely) worst-case $A_{\rm drag}$  $\approx 10^{-11}$ m s$^{-2}$; it turns out that \eqi \left\langle\Delta
  N_{\rm drag}\right\rangle_{T}\sim 1\times 10^{-5}\ {\rm m}.\lb{drr}\eqf But $A_{\rm drag}$ is not constant over time spans days or years long \cite{For06}, so that such an effect is not a concern here.  By the way, even if it was not so, by assuming a $\approx 10\%$ mismodeling in
  drag--which is, in fact, modeled by Konopliv et al. in Ref. \cite{Kon06}--\rfr{drr}, mapped onto about 5 yr, would give a $\approx 0.7\%$ uncertainty.

  Finally, Krogh in Ref. \cite{Kro07} remarks that decreasing in the averages of the RMS orbit overlaps occurred in view of constantly improved modeling \cite{Lem01,Yua01}, but he does not recognize that the improved modeling of the non-gravitational forces acting on MGS introduced in
  Ref. \cite{Kon06}
with respect to
previous works \cite{Lem01,Yua01} in which the Lense-Thirring effect was not modelled as well,
{\it only affected in a relevant way just the along-track} RMS overlap differences (a factor 10 better than in Ref. \cite{Lem01,Yua01}),
{\it not the normal ones} (just a factor 2 better than in Ref. \cite{Lem01,Yua01}).
Moreover,
if the relativistic signature was removed or not present at all so that the determined
out-of-plane RMS overlap differences were only (or mainly) due to other causes
like mismodeling or unmodeling in the non-gravitational forces,
it is difficult to understand why the along-track RMS overlap differences
(middle panel of Figure 3 of Ref. \cite{Kon06}) have almost the same
magnitude, since the along-track component of the MGS orbit is
much more affected by the non-gravitational accelerations (e.g.
the atmospheric drag) than the out-of-plane one.

         \end{itemize}

\section*{Acknowledgments}
I gratefully thank A. Konopliv, NASA Jet Propulsion Laboratory
(JPL), for having kindly provided me with the entire MGS data set.
 I am indebted  to E. Lisi, P.
Colangelo, G. Fogli, S. Stramaglia, and N. Cufaro-Petroni (INFN,
Bari) for very stimulating discussions.


\end{document}